\documentclass[a4paper]{jpconf}
\usepackage{graphicx}
\begin{document}
\title{Linearized force constants method for 
lattice dynamics in mixed semiconductors}
\author{Ayoub Nassour, Joseph Hugel, and A V Postnikov}
\address{Laboratoire de Physique et des Milieux Denses,
Institut de Physique \'electronique et Chimie,
Paul Verlaine Universit\'e--Metz, 1 Boulevard Arago, F-57078,
Metz  Cedex 3, France}
\ead{postnikov@univ-metz.fr}
\begin{abstract}
A simple and accurate method of calculating phonon spectra in mixed 
semiconductors alloys, on the basis of preliminarily (from first principles) 
relaxed atomic structure, is proposed and tested for (Zn,Be)Se and (Ga,In)As 
solid solutions. The method uses an observation that the interatomic force 
constants, calculated \emph{ab initio} for a number of microscopic 
configurations in the systems cited, show a clear linear variation of the main
(diagonal) values of the interatomic force constants with the corresponding 
bond length. We formulate simple rules about how to recover the individual 
3$\times$3 subblocks of the force constants matrix in their local 
(bonds-related) coordinate systems and how to transform them into a global 
(crystal cell-related) coordinate system. Test calculations done for 
64-atom supercells representing different concentrations of (Zn,Be)Se and 
(Ga,In)As show that the phonon frequencies and compositions of eigenvectors 
are faithfully reproduced in a linearized force constants calculation, 
as compared to true \emph{ab initio} calculations.
\end{abstract}
Calculation of phonon spectra in the harmonic approximation 
(frequencies $\omega$, eigenvectors $A^{\beta}_k$)
becomes straightforward from a simple diagonalization,
\begin{equation}
\sum_{\beta,k} \left[\,
\omega^2 \delta_{\alpha\beta}\delta_{ik} -
\frac{D^{\alpha\beta}_{ik}}{\sqrt{M_{\alpha}M_{\beta}}}\right] A^{\beta}_k = 0
\label{eq:dynam}
\end{equation}
($\beta$ runs over atoms in the unit cell,
$k$ over Cartesian directions)
once the matrix of force constants,
\[
D^{\alpha\beta}_{ik} = \frac{\partial^2 E_{\rm tot}}{%
\partial X^{\alpha}_i \,\partial X^{\beta}_k} = D^{\beta\alpha}_{ki}
\]
is known. In the adiabatic Born--Oppenheimer approximation, the forces 
and their derivatives over small atomic displacements can be well yielded 
by a density functional theory (DFT).

The calculation approaches, in their turn, are either of linear-response
type (see Ref.~\cite{RMP73-515} for a review), or of 
frozen-phonon \cite{PRB24-2311} or otherwise finite-displacement 
type -- see, e.g., Ref.~\cite{PRL78-4063}.
All \emph{ab initio} methods work reasonably well for perfect crystals
and not so large supercells. The situation becomes more difficult
if dynamic properties of alloys must be analyzed. In the present
contribution, we deal with substitutionally disordered II-VI and III-V 
semiconductor alloys A$_x$B$_{1-x}$C in the absence of structural defects, 
so that the topological structure of cation/anion sublattices in the zincblende
phase is preserved. However, chemical substitutions on either cation or anion
sublattice yield considerable displacements of ions from perfect 
crystallographic positions. It is known 
since early 1980s \cite{PRL49-1412} that the average lattice constant 
in a semiconductor alloy follows Vegard's law whereas individual A-C, B-C 
anion-cation bond lengths tend to maintain their values as in parent binary 
compounds. A complex relaxation pattern appears from an interplay of these
two tendencies in each given microscopic configuration.
The variations of bondlengths according to local environment are up to 
several per cent, and may give rise to well resolved Raman lines. 
In fact, the Raman spectroscopy can serve as a tool
for probing these local environments, that would ideally require  
a reference to model first-principle calculations, imitating different
environments. An example of such treatment for (Zn,Be)Se alloys in the context
of elucidating the effects of structural ``percolation'' is given
in Ref.~\cite{PRB-ZBS}, and in the context of partial ordering
tendencies in the same alloy -- in Ref.~\cite{ZnBeSe_ord}.
The lattice relaxation (which 
requires forces on atoms to be available in a calculation) is by far
easier than the subsequent calculation of force constants by 
linear response or finite-displacement method. In order to overcome
this bottleneck, we propose in the present contribution a fast
and simple way to estimate the force constants in any given atomic
configuration of a pseudobinary semiconductor alloy. Our approach 
is based on clear trends resulting from a number of first-principle 
calculations, and does not need empirical parameters to match 
the experimental data.

%
Our two basic assumptions are the following: \\
$i$) the interatomic interactions in the force constant matrix of
a semiconductor alloy can be limited by the nearest neighbours
(NN; cation--anion) and next-nearest neighbours (NNN; cation-cation
and anion--anion)
terms and discarded for more distant pairs, at least for the sake
of obtaining the vibration spectra with accuracy typical for the
Raman spectra of these crystals; \\
$ii$) the NN and NNN blocks of the force constants matrix
can be recovered from  the corresponding interatomic distances and
angles. The actual prescription for this follows from fully
\emph{ab initio} (finite-displacements) phonon calculations done
by the {\sc Siesta} method\cite{siesta}.

\begin{figure}[b]
\centerline{\includegraphics[width=0.96\textwidth]{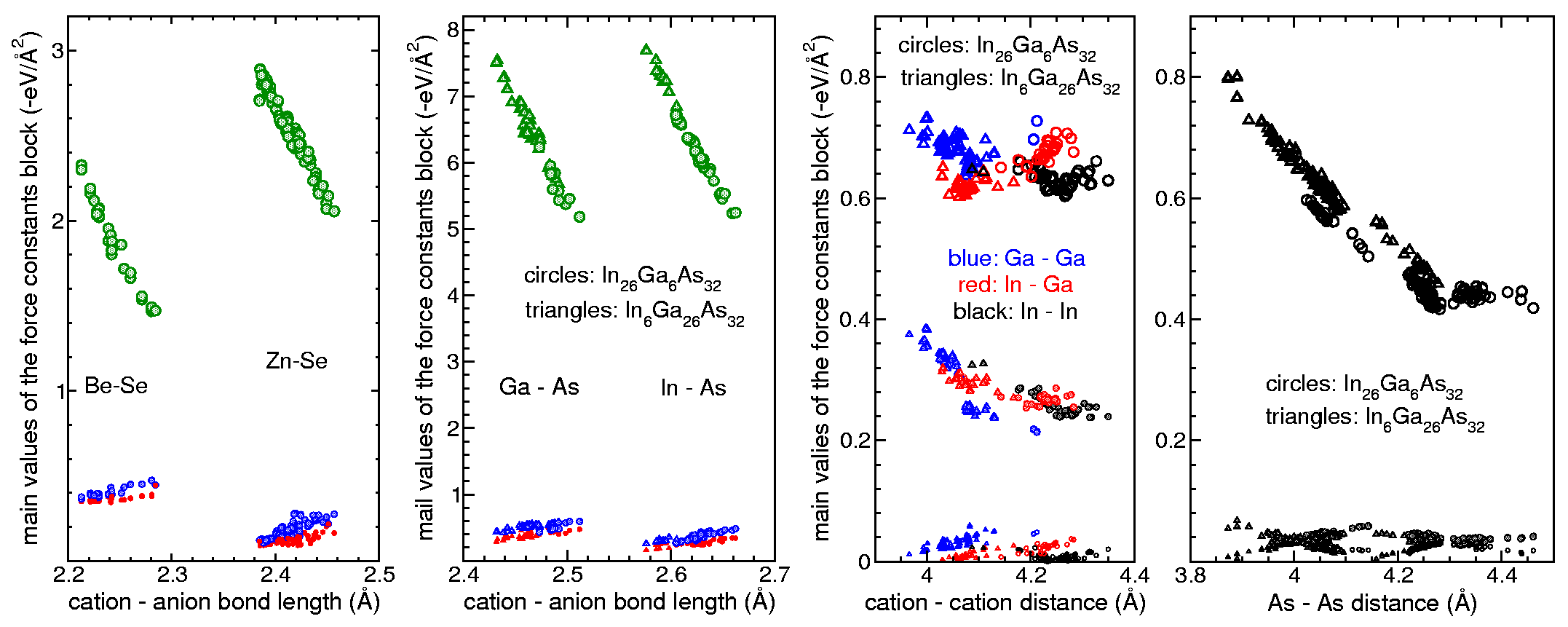}}
\caption{\label{fig:FC}
Main elements of the force constants blocks relating NN and NNN in
(Zn,Be)Se and (Ga,In)As alloys. The data are collected over a number
of 64-atom supercells with different concentration. The NNN interactions
in the (Zn,Be)Se system were shown in Fig.~6 of Ref.~\cite{PRB-ZBS}.
}
\end{figure}

The supercells used
were a prototype ``chain+impurities'' Be$_6$Zn$_{26}$Se$_{32}$
one described in Ref.~\cite{PRB-ZBS}, ``Be-clustering'' supercells
of the same size containing 1$\dots$4 Be atoms grouping around the same 
Se atom in ZnSe; the same supercells constructed for (Ga,In)As system,
and a number of smaller (32-atom) supercells for both alloys, which were
subject to discussion in Ref.~\cite{ZnBeSe_ord}. 
The phonon densities of states (PhDOS) of species $\alpha$ discussed 
in the following are zone-center-projected components of eigenvectors, 
sampled as
$$
n_{\left\{\alpha\right\}}(\omega,{\bf q}) = 
\sum\limits_{\nu} 
\delta(\omega - \omega_{\nu})
\sum\limits_i 
\Bigl| 
\sum\limits_{\alpha} A^{\alpha}_i(\omega_{\nu})\, e^{i{\bf q}{\bf R}_{\alpha}} 
\Bigr|^2
$$
(${\bf q}$=0 in our case), and technically 
obtained from zone-center discrete vibration spectra $\omega_{\nu}$
in various supercells, artificially broadened with half-width parameter
of 10 cm$^{-1}$.  

The first of the above statements is verified by comparing the PhDOS
calculated with the elements of $D^{\alpha\beta}$ set to zero beyond
the second neighbours. The result was found,
for any practical purpose, indistinguishable from that calculated
with the full force constants matrix. On the contrary, an omission
of NNN interactions leads to unacceptable damage of the PhDOS.

\begin{figure}[b]
\parbox[b]{0.4\textwidth}{
\centerline{\includegraphics[width=0.3\textwidth]{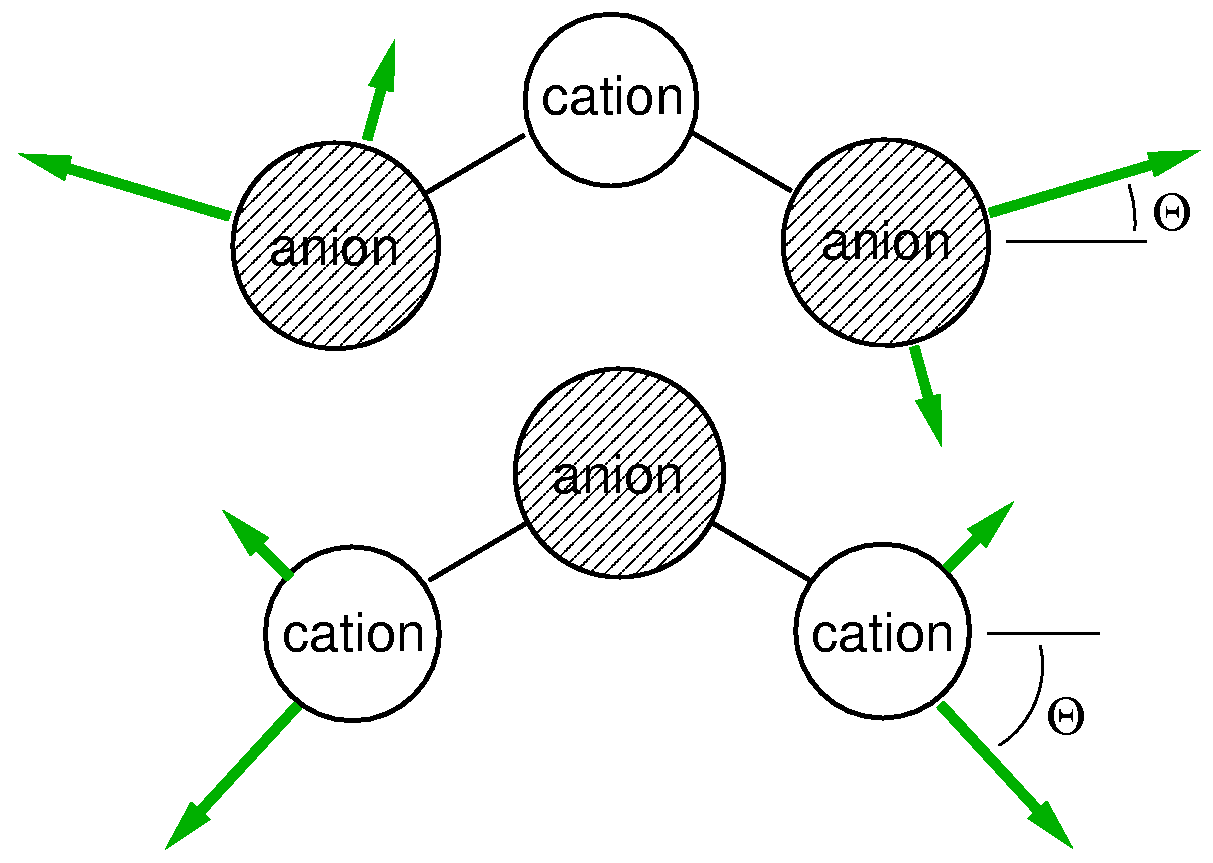}}
}
\parbox[b]{0.6\textwidth}{\caption{\label{fig:NNNaxes}
The orientation of local axes (in which the two-atom block
of the force constant matrix is diagonalized) for cation--cation
and anion--anion NNN. 
Note that $\Theta<0$ in the first case and $>0$ in the second.
See text for details.
}}
\end{figure}

In order to understand the behaviour of NN and NNN force constants,
we diagonalized them blockwise, for each given pair of atoms involved.
More specifically, $D^{\alpha\beta}$ and $D^{\beta\alpha}$ blocks
of the global force constant matrix, both having dimension 3$\times$3,
can be cut out of the global matrix to form a (symmetric)
6$\times$6 block. Its diagonalization yields antisymmetric (three
positive and three negative) values. 
Let us discuss first the diagonal elements of the force constants
matrix for the NN. The major element comes from the central force,
acting along the corresponding bond, and two minor elements from tangential
forces of markedly similar size, revealing 
an essentially axial anisotropy of NN interaction with respect to
the bond direction. 
Fig.~\ref{fig:FC} shows diagonalized force constants for (Zn,Be)Se
alloys (those previously shown in Ref.~\cite{PRB-ZBS}) along with new results 
for (Ga,In)As. We address the reader to  Ref.~\cite{PRB-ZBS} for a discussion 
on a more ionic character of Zn--Se bonds in contrast to more covalent
character of the Be--Se bonds which follows from these results;
obviously there is no such strong distinction in the degree of
covalence between Ga--As and In--As bonds. 
What is striking for
both mixed systems is a markedly linear variation of the force constant
values with the bond length. We emphasize that the data shown
in Fig.~\ref{fig:FC} are collected over a number of supercells
with different concentrations, hence this trend must be quite independent
on the details of microscopic structure.
We approximated this dependence by a linear function,
with different coefficients for each given cation-anion pair.
With the diagonal elements coming from the fit, one has yet
to specify the local system in which the force constant matrix
is reduced to its main axes. Analyzing the transformation matrices
(those which yield diagonalization) over a number of systems, we conclude
that the main axis should indeed be directed along the bond,
and the orientation of two others is in fact irrelevant because of
the abovementioned axial symmetry; we arbitrarily set 
the second axis in the plane passing through the cation--anion pair and 
whatever the closest third atom to any of them.

\begin{figure}
\centerline{\includegraphics[width=0.8\textwidth]{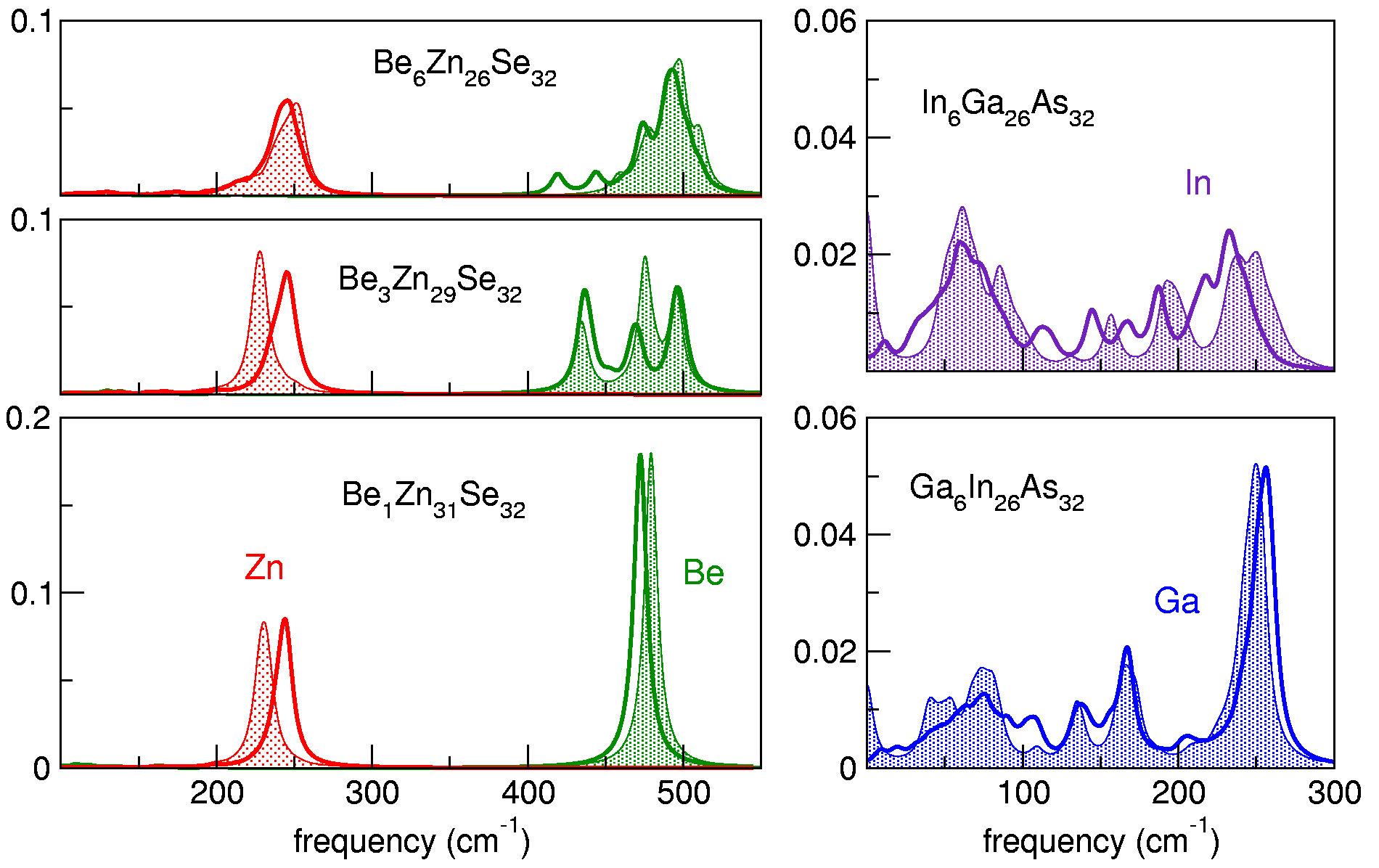}}
\caption{\label{fig:phdos}
Zone-center projection of the local PhDOS
in several supercells of (Zn,Be)Se and (Ga,In)As composition.
Shaded areas: fully \emph{ab initio}, thick lines: 
with linearized force constants. 
}
\end{figure}

For NNN interactions, the orientation of the local system which
diagonalizes the force constants matrix is not that straightforward.
Our analysis shows that the main axis of cation--cation or
anion--anion interaction lies in the plane which passes through
these two atoms and the third one (anion or cation, respectively)
which is bonded to the two in question; see Fig.~\ref{fig:NNNaxes}. 
The direction of this main axis in \emph{not} along the line connecting the NNN
in question, but forms with it an angle $\Theta$, practically unaffected  
by the NNN distance, but different for different types of NNN.
The angles $\Theta$ for (Zn,Be)Se systems are
$-47^{\circ}$ (Be-Be and Be-Zn),
$-39^{\circ}$ (Zn-Zn),
$+16^{\circ}$ (Se-Se);
for (Ga,In)As system --
$-52^{\circ}$ (Ga-Ga),
$-41^{\circ}$ (In-In),
$-46^{\circ}$ (In-Ga),
$+28^{\circ}$ (As-As).
The second (in the size of the corresponding element) main axis is also
in the plane connecting three atoms and normal to the first,
and the third one (not shown in Fig.~\ref{fig:NNNaxes})
is normal to the plane. 

It rests to specify the on-site $D^{\alpha\alpha}$ blocks
of the force constant matrix, which are clearly dominant and strongly
dependent on each atom's environment in crystal. We recovered these blocks
from the acoustic sum rule,
$
\sum_{\beta} D^{\alpha\beta}_{ij} = 0 = \sum_{\alpha} D^{\alpha\beta}_{ij}\,,
$
assuming $D^{\alpha\beta}$=0 beyond NNN.

In conclusion, we test the results of the phonon spectrum calculations
with the parametrized force constants, against those obtained
with the ``true'' force constants from \emph{ab initio} calculations
for the corresponding supercells. The PhDOS shown in Fig.~\ref{fig:phdos}
show, to our opinion, a quite satisfactory level of agreement,
which is about the same for all supercells we have considered.
Based on this, we suggest that the present approach of 
``linearized force constants'' may be useful for fast calculations
of vibration spectra in large supercells. 

The present contribution serves the justification and description
of our approach. New results obtained with
this method for larger supercells, along with the parametrizations for
other mixed semiconductor systems we have studied, will be reported elsewhere.

\ack
The use of computational facilities at the Centre Informatique
National d'Enseignement Sup\'erieur (CINES project N$^{\circ}$pli2623,
Montpellier, France) is greatly appreciated.

\section*{References}
\bibliographystyle{iopart-num}

\providecommand{\newblock}{}

\end{document}